# Public Intervention Strategies for Distressed Communities

**Lester O. King, PhD.**
**Rice University,**
**Houston, TX, USA**
kingL2@gmail.com

**ABSTRACT**
*This research presents a methodology to comprehensively define Distressed Communities. We further identify if there is a significant difference in public investment between Distressed communities and Wealthy communities.*

*One of the key tools in sustainability planning is the use of sustainability indicators (SIs). A considerable amount of scholarship has contributed to define and develop SI programs for local level application (Elgert & Krueger, 2012). Much of the focus of SI research is on developing the ideal indicator based on defined criteria for each indicator (Hart, 1999; Innes & Booher, 2000; Holman, 2009). Here we suggest a methodology beyond defining the ideal indicators to demonstrating how indicators can be used for more in depth analysis of complex urban problems.*

*In this analysis we reduce 34 development metrics to a smaller number of factors that represent how the data can be classified into groups based on similarities among 88 communities. Using the factor (group) that contained measures identifying Distressed Communities, the communities were allotted an index score and ranked. The top 10 communities were then compared to the bottom 10 communities according to 14 place based variables related to opportunities for local government led improvement.*

## INTRODUCTION
Planning in the US is still forging new ways to integrate sustainable development, which is a comprehensive mantra for development planning (Berke and Manta-Conroy 2000; Jepson & Haines, 2014). One of the key tools in sustainability planning is the use of sustainability indicators (SI). A considerable amount of scholarship has contributed to define and develop SI programs for local level application (Elgert & Krueger, 2012). Much of the focus of SI research to-date is on developing the ideal indicator based on defined criteria for each indicator (Hart 1999; Innes and Booher 2000; Holman, 2009). This paper furthers the dialogue by offering a methodology beyond defining the ideal indicators to demonstrating how indicators can be used for more in depth analysis of complex urban areas.

Using data to drive public policy can be proven to be beneficial through explicit inclusion in policy (Innes, 1988; Brugmann 1997; Bell and Morse 2008). Sustainability indicators can go beyond the singular function of measuring gains or losses in public policy to informing and enhancing the framing of public policy. This is because SIs represent a very value laden body of social, economic and environmental determinism in development services. SIs represent the theory of sustainable development, which is composed of a definition, set of principles, frameworks, suggested development processes and indicator characteristics. SIs can be thus used to serve political and operational functions. On the operational gamut, they are useful for goal oriented objectives (Backhaus, Bock, & Weiers, 2002). This is why it is important to further demonstrate how indicators can be applied towards more in-depth research to tackle complex problems in urban areas.

The research first identifies a set of indicators developed with the assistance of a group of local experts in Houston, Texas. Data was collected and analyzed to measure the indicators at the level of 88 communities in the city. Exploratory analysis was conducted on the data collected for each community to determine how the indicators were related and to identify any relationships in the data that were meaningful findings for focusing planning policy. Specifically we were interested in answering the question of 'What are the factors that can best identify and explain distressed communities in Houston?' After the most appropriate factor to explain distressed communities was identified, scores were developed for each community according to the factors and the communities were ranked according to level of distress. Next the top ten most distressed and bottom ten least distressed were compared only according to the indicators from the dataset that explain place based differences. Specifically differences for which public intervention may support better quality of life and improvements to community function. The two groups were compared to answer the question, 'Is there a significant difference between the most distressed and best performing communities in Houston according to community characteristics?' The paper concludes with policy recommendations for public improvements to distressed communities in Houston. This research demonstrates the use of performance measurement as an essential component of planning in general and then highlights the use of sustainability indicators in particular as a robust, comprehensive and systematic approach for assessment and enhancement of planning policy.

**BACKGROUND**

In the 1990s as strategic thinking and results oriented management was improving in the public sector, the use of measurement systems for tracking performance became widespread in the US. The book, 'Reinventing Government', published in 1992 was a landmark guide for local government performance management. The federal government also implemented the 1993 Government Performance and Results Act to further spur this movement (United States Congress, 1993).

At the same time that US government agencies were becoming more efficient through the implementation of performance measurement; crises in planning were being reported in terms of a loss of unified vision and inability to meet stakeholder needs (Levy, 1992). A review of typologies of plans conducted in 1995 showed that the *Development Plan* was the closest instrument to actually having performance measures through its structure of linking goals, objectives and policies. This was exemplified by the Sanibel, Florida plan. However none of the plans presented in the *Stalwart Family Tree* explicitly incorporated performance measures (Kaiser & Godschalk, 1995). This gap was later identified as the need to integrate performance measures to link plan goals with actual outcomes (Talen, 1996; Berke & Manta-Conroy, 2000). This national movement in the 1990s to develop performance measurement was actually coupled with an international movement to develop sustainability indicators to measure development perfromance (United Nations, 1996). Talen (1996) and later, Berke and Manta-Conroy (2000), show that these movements did not go unoticed by the academic community. Articles written between then and now highlight the gradual establishment of indicator projects across cities and countries abound (Mori & Christodoulou, 2012). However, these projects are not explicitly linked to plans and in most cases they are conducted outside the planning department. What is even more puzzling, is that even though many plans have attemted to incorporate sustainability, and one of the major instruments of incorporating sustainability is to develop sustainability indicators, indicators are still not normatively included in most plans developed[1]. (Berke & Manta-Conroy, 2000; Conroy & Beatley, 2007).

The American Planning Association (APA) recently assembled a task force to establish a standard for developing comprehensive plans which meet the intent of sustainable development (APA, 2010). One of the implementation suggestions was to use performance metrics to monitor implementation progress. This essential multi-year effort by the APA represents a desperately needed stance on the establishment of what should constitute a good comprehensive plan. The elements are however described and presented without performance metrics, which leaves the door wide open for local areas to demonstrate various levels of implementation, without clear guidance on what would be appropriate. In contrast the United States Green Building Council (USGBC), under its Leadership in Energy and Environmental Design (LEED) rating system, has established clear performance metrics for development projects to meet various levels of performance (USGBC, 2009).

We have a gap in planning in terms of developing performance metrics for plans. Talen (1996), presented an excellent overview of established approaches to developing performance metrics in the field of planning. She succintly summarized the problems by recognizing the difference between *planning* implementation and *plan* implementation (Ibid. pg 249). In planning implementation we have several models for evaluating alternative plans and planning documents. Methods for evaluating planning behavior and planning impacts and also methods for implementation analysis. What we do not have are many methods for evaluation of implementation of *plans*. The problem was attributed to three concerns (1) lack of clarity with defining success; (2) lack of methodologies for addressing multicausality due to the complexity of living environments; and (3) lack of suitable quantitaive methods attributed to missing data, statistical validity and ability to account for the subjectivity of reality.

Sustainability indicators can help to bridge this gap based on the fact, that they are not just performance measures, but are value laden descriptors and guides for helping to formulate planning and development goals. Sustainability Indicators can easily serve monitoring, evaluation or steering functions (Wiek & Binder, 2005). Sustainability has established the precept of addressing the build environment from the three pillars of social, economic and environmental development. Sustainability planning research has also established that not doing so means that we cannot properly address critical conflicts that exist and are difficult to frame such as devleopment conflicts and resource conflicts (Campbell, 1996). Sustainability ensures that we strive to identify connections between disparate systems for addressing trade-offs (Neuman & Churchill, 2011).

To address Talen's (1996) problems with establishing clarity for defining success, sustainability indicators can help if the development of indicators completes the initial component of plan preparation, which is to develop goals. In other words goals and objectives are developed and then indicators identified to measure those objectives. The indicators will constitute the tools for an essential systematic evaluation, through which we can re-evaluate planning elements (Berke & Godschalk, 2009). In practice we develop goals based on normative thinking and then develop plan elements to address those goals without recognition of agreed upon standards to measure success. Therefore our plans have little intellectual or political mandate (Altshuler, 1965).

---

[1] One exception is the San Deigo Association of Governments, which established a performance monitoring program for its regional comprehensive plan in 2006 (SANDAG, 2006).

Integrating sustainable development indicators into comprehensive planning would help to solve some of our problems. But we also have the issue of multiple stakeholders and departments oftimes having mandate for planning different thematic areas, such as a parks department or a housing department. Kaza & Hopkins (2012) also point out the difficulty of managing intentions and actions from multiple stakeholders in plans. They suggest that we would have to recognize spatial, temporal and functional relationships between actions to properly address the complexity of the built environment.

A focus on understanding how SIs can bridge the gap between disparate functions in city management and planning can help planning departments better coordinate with other departments in meaningful ways. Coordinating multiple city departments to understanding the factors that drive crime in an urban area, for example, should not mean the collection of separate topical plans to create a consolidated plan. This process is best done by aggregating indicators identified by different experts within different departtments, such as transportation, housing, and economic devleopment. Then analysis is applied to find predictors for increased crime. The improvement or reduction in these predictors/ indicators are systematically formed and reliable assumptions that could drive the development of a robust crime reduction plan. Thus a study of an in-depth application of SIs, such as this one, can highlight the benefits of including SIs in planning.

**Managing complex planning problems without planning**

The city of Houston is an example of a major city in the US (4th Largest), where complex issues arise that need complex solutions on a day to day level. The City of Houston, Texas is the 4th most populous city in the United States with 2,099,551 persons in 2010 (US Census 2010). Based on its rate of growth, calculated since 1980, the population is projected to grow to 3,172,082 by 2060. This is adding a little over 50% more people or just over 1 million. Houston is the most sprawling city of the top 4 most populous cities in the US, with a land area of approximately 600 square miles, which is larger than the land area of New York and Chicago combined. According to the aforementioned population projection, in 2040 Houston will still be less dense than New York, Chicago, or L.A. are today.

|  | Houston 2010 | Houston 2040 | New York | L.A. | Chicago |
|---|---|---|---|---|---|
| Population | 2,099,451 | 3,172,082 | 8,175,133 | 3,792,621 | 2,695,598 |
| Land Area 2010 (sq.ml) | 600 | 600 | 303 | 469 | 228 |
| Population Density | 350,142 | 529,033 | 2,701,630 | 809,179 | 1,184,358 |

Houston is a minority- majority city leading the way to what we can expect for the entire US in the near future in terms of demographics (Economist, 2009). Houston has 44% Hispanic, 26% White, 23% African American and 7% of the population from other racial/ethnic groups. As of the 2010 census the White group was 300,000 persons smaller and the Hispanic group was 640, 000 persons larger than compared to 1980 numbers ( U.S. Census Bureau, 2011). Houston has the highest violent and property crime rates among the largest cities in the country (US Census, 2007). The poverty rate is 23% and the median household income is $42,355. The average household spent 46% of their income on housing and transportation costs (King, 2012). This city is going through a tremendous demographic change; crime and poverty are high. Incomes are low on average and people spend a tremendous amount of their income on housing and transportation costs. It is not clear that the average Houstonian understands these major issues since popular media primarily focuses on housing prices without regard for incomes and or focuses on job numbers without regard for the pay scale of those

jobs or the fact that the job numbers are for the Houston metro area and not just for the City of Houston (Brennan, 2012; Mulvaney, 2014).

Houston is a good case study for this topic because it has no comprehensive plan and has several stakeholders producing various forms of plans. In fact, the city of Houston, has not established goals, or visions to direct development in the city. Therefore existing development should only be explained through separate and distinct projects and thematic investments. A report has identified 35 separate and distinct plans extant in the city (HNTB & Spillette, 2003). Therefore we can analyze development in Houston as a lab neutral to central planning to determine if patterns can be identified which support accepted planning dogma. We can also determine the strength of indicators to identify areas in which development performance can be improved based on current or suggested governance structures in the city.

Planning without the legal basis for planning is important to the city as can be demonstrated by the recently developed voluntary development review process for municipal management districts (MMDs) (Gafrick, 2013). There are now 19 active MMDs in Houston covering approximately 30% of land area. These districts have combined operating budgets of $42M, although they very much act independently. These MMDs were created by state legislature, with the city's only oversight being at inception with the decision of whether or not to approve the creation of the MMDs.

**Data Sources**
Data for this analysis comes from the Houston Sustainability Indicators Project (HSI) at the Shell Center for Sustainability on the campus of Rice University. Thirty four (34) measures representing twenty two (22) indicators are included in the analysis. The data represents a balance of social, economic and environmental measures. The data includes: (1) U.S. Census demographic data such as race/ethnicity, income, housing value; (2) Street network centerlines, street condition assessment, storm sewer assessments, air pollution, jobs, bus stops, and park locations all in GIS shapefile vector data from various local, regional and state government departments in Houston. (3) GIS shapefile raster data on land cover from U.S. Geological Survey (4) GIS shapefile vector data for political outlines from the local planning department.

**METHODOLOGY**
In the analysis that follows we reduce 34 measures of development to a smaller number of factors that represent how the data can be classified into related groups based on similar trends among 88 communities in Houston.

Using the factor (group) that most contained measures identifying distressed communities, the communities were allotted an index score and ranked.

The top 10 communities were then compared to the bottom 10 communities according to 14 place based variables related to opportunities for public improvement.

**Houston Sustainability Indicators**
Data used in this study comes from the Houston Sustainability Indicators program (HSI). The HSI was developed on the campus of Rice University to assist with the characterization of sustainable development in Houston and to support the long term monitoring of development performance. Now in its fourth year, the program has published three SI reports looking at city level performance over time and projected into the future and also at the performance of major districts within the city (SCS, 2013).

The HSI utilizes the Theme /Sub-Theme (TST) framework and has identified themes and sub-themes based on the 'Big Ideas' that Houstonians rally around. These 'Big Ideas' were identified from various sources such as a popular annual survey of opinions and aspirations of Houstonians[2]; local media articles; and discussions and meetings with several experts and citizens throughout the city. Indicators were selected based on the definition of sustainability, principles, framework, criteria and functions presented in this paper. Data was collected and analyzed from various sources ranging from federal, state, and regional to local agencies. Experts were convened among several workshops to review the data, assign ratings for trends towards sustainability, and suggest policy to improve the indicators. Reports were then developed illustrating and discussing the findings, presenting background relevance to sustainability of each indicator and presenting policy options for improving the indicators.

The HSI reports have achieved major public impact with feature stories on the cover of the leading local newspapers (Sarnoff, 2013); feature interviews on major local television stations; headline news features on local radio outlets such as National Public Radio (NPR); and consultation requests from elected officials, business executives, and local non-profits. This public impact in itself hints to the value of the HSI program to supporting sustainable development in Houston.

The goal for the HSI project was to capture development themes based on recognition of the various levels of governance overlapping the city and highlighting the issues that may be found at various levels of geography such as neighborhoods, regions, urban and sub-urban dynamics. The first two reports published by the center[3] were focused on the Houston municipal boundary. The third report focused on the eleven council districts within the city, measuring performance at the levels of elected officials[4]. A fourth report is compiled at a lower level of geography, which is the community level[5].

The following table presents the indicators selected for the HSI program.

Table 3. Houston Sustainability Indicators

| Social Indicators | | |
|---|---|---|
| **Theme** | **SubTheme** | **Indicator** |
| Social Demography | Population Growth | Population Growth |
| | Education | Education Attainment |
| | Community Involvement | Voter Participation |
| Poverty | Inequality | Income Inequality |
| | Poverty Level | Poverty Rate |
| | Healthcare Delivery | Health Coverage |
| Livability | Cost of Living | Affordability |
| | Quality of Life | Accessibility of Public Spaces |
| | Health & Nutrition | Food Deserts |
| **Economic Indicators** | | |
| **Theme** | **SubTheme** | **Indicator** |
| Economic Development | Employment | Employment Status |
| | Macroeconomic Performance | Primary Jobs/ Green Jobs |

---

[2] (Klineberg, 2010)
[3] (Blackburn 2011; King 2012)
[4] (King, 2013)
[5] (King, 2014)

|  | Earnings | Income |
|---|---|---|
|  | Waste Generation & Management | Waste Generation |
|  | Energy Use | Energy Consumption |
| Transportation | Access | Access to Public Transportation |
|  | Demand | Vehicle Miles Travelled |
|  | Mode | Travel Choice |

**Environmental Indicators**

| Theme | SubTheme | Indicator |
|---|---|---|
| Atmosphere | Air Quality | Ambient Pollutants |
|  | Climate Change | Greenhouse Gas Emissions |
| Fresh Water | Water Quality | Water Pollution |
|  | Water Demand | Water Use |
|  | Water Resources | Water Availability |
| Land | Flooding | Floodplain Expansion |
|  | Land Cover | Land Cover Change |
|  | Classification | Jobs/ Housing Balance |
|  |  |  |

## ANALYSIS
### Principal Components Analysis
Factor analysis was used to reduce the many variables to a smaller set representing important clusters based on data trends across all of the communities. This methodology reveals clusters of data when measures exhibit similar trends. Principal components analysis was used to extract all the individual differences between indicators. Eigenvalues over 1 was used to determine which factors were chosen based on the amount of variance explained by a given factor. Orthogonal rotation was selected since the assumption is that the factors are uncorrelated. The varimax method was selected for orthogonal rotation since this method further forces distinctions between the factors (Frear 2007). All communalities returned were greater than .5 indicating that the resulting factors explained the variances of the indicators very well (Sorensen 2007). Nine (9) factors were extracted with eigenvalue >1. The percent of variance was 77.5% of the total variance. The final overall KMO was middling at 0.707 (Kaiser1974), but this was enough to run factor analysis although it shows that observed correlations between variables are not completely explained by other variables (Norusis 2006).

### Factor Scores
To determine the score for a neighborhood on a particular factor, the neighborhood's data on each variable is multiplied by the factor weight for that variable. The sum of these weight-times-data products for all the variables yields the factor score. Neighborhoods will have high or low factor scores as their values are high or low on the variables entering a pattern. Factor score, case j and factor k

$$F_{jk} = \sum_{i=1}^{p} W_{ji} X_{ik} \quad (1)$$

Where *F* - factor score; *j* – case; *k* – factor; *W* – factor score coefficient; *X* – factor; *p* – number of variables

| Variable | Factor 1 Distressed Communities | Factor 2 Inner City | Factor 3 Growth Areas | Factor 4 Minorities | Factor 5 Ethnic Areas | Factor 6 Industrial | Factor 7 Single Use | Factor 8 Flood Zone | Factor 9 Adequate Sewers |
|---|---|---|---|---|---|---|---|---|---|
| MedianIncome | **.947** | .054 | .100 | .056 | .081 | -.062 | .016 | .020 | .040 |
| HealthSpending | **.934** | .040 | .046 | .185 | .070 | .008 | -.014 | .071 | .018 |
| %BelowPoverty | **-.893** | .118 | -.073 | .071 | .058 | -.004 | -.119 | .017 | -.151 |
| H+Tcosts | **.871** | -.314 | .017 | .098 | .005 | .026 | .078 | .095 | -.009 |
| MedianValuehouse | **.834** | .324 | -.037 | .193 | -.080 | .047 | .007 | -.030 | .052 |
| %White | **.831** | -.023 | .109 | -.066 | .013 | -.152 | .016 | .161 | -.333 |
| %MastersDegrees | **.802** | .366 | .065 | .323 | -.052 | -.091 | .086 | -.051 | -.054 |
| %Unemployed | **-.690** | -.297 | -.137 | .203 | .251 | -.072 | -.098 | .001 | -.072 |
| %TransitRiders | **-.534** | .406 | .011 | .505 | -.109 | -.122 | -.106 | -.001 | -.247 |
| VMT | -.133 | **-.889** | .027 | -.239 | .177 | -.026 | -.041 | .165 | .016 |
| %Persons1/4MiletoBusStop | -.108 | **.826** | -.067 | .324 | -.057 | .228 | -.066 | -.073 | -.050 |
| %OpenSpace | -.044 | **-.772** | -.018 | .039 | -.146 | .004 | .135 | -.147 | .213 |
| IntersectionsPerSqMile | -.011 | **.738** | -.138 | .108 | .383 | .051 | .053 | -.018 | -.129 |
| %PersonsinFoodDesert | -.110 | **-.722** | -.073 | .075 | .177 | -.117 | .014 | -.041 | -.134 |
| %HighIntenseDevelopment | -.137 | **.716** | .043 | -.223 | -.241 | -.252 | -.157 | -.298 | -.109 |
| DistancetoCBD | .226 | **-.671** | .440 | -.217 | -.131 | -.127 | .250 | .175 | .034 |
| %HousinginBusinessCenters | .404 | **.635** | -.030 | -.004 | -.143 | -.124 | .123 | -.257 | -.063 |
| %Persons1/4MiletoPark | .155 | **.617** | -.163 | .076 | .197 | .049 | .007 | .453 | .097 |
| AverageWaterUse | .208 | .052 | **.867** | .083 | .057 | .095 | .088 | .015 | .160 |
| PopulationGrowth90-10 | .190 | -.203 | **.815** | -.033 | .009 | -.184 | -.029 | -.021 | .122 |
| PopulationDensity | -.026 | -.023 | **.809** | -.017 | .078 | .164 | -.102 | .102 | -.131 |
| Persons1/4MiletoWasteSite | -.207 | .064 | **.397** | -.093 | -.152 | -.167 | -.048 | -.272 | .297 |
| %Hispanics | -.305 | -.240 | -.107 | **-.783** | .042 | .075 | -.154 | .010 | -.041 |
| %Voters | .336 | .090 | -.122 | **.688** | .156 | .340 | -.157 | .130 | .066 |
| %Black | -.455 | -.421 | -.112 | **.545** | .129 | -.116 | .184 | -.133 | -.308 |
| %PrimaryJobs | -.033 | .083 | -.225 | -.009 | **-.797** | .013 | -.154 | .053 | .009 |
| %OtherRace | .413 | -.049 | .267 | -.092 | **-.469** | -.166 | .232 | -.008 | -.392 |
| #AirExceedances | -.073 | -.022 | .100 | -.115 | -.037 | **.764** | .014 | -.188 | -.200 |
| %Low-MidIntensityDevelopment | -.047 | .348 | -.086 | .311 | .047 | **.709** | .005 | .160 | .283 |
| LandDiversity | .028 | -.227 | -.223 | .014 | .120 | -.042 | **.767** | .093 | .053 |
| %PoorStreets | .207 | .446 | .182 | .098 | .087 | .069 | **.535** | .071 | .201 |
| %Spending>30%IncomeonHousing | -.182 | .066 | -.206 | .000 | .456 | -.071 | **-.485** | -.086 | .157 |
| %PersonsinFloodZone | .052 | -.158 | .108 | -.005 | -.147 | -.123 | .145 | **.864** | -.035 |
| %AdequateSewers | .105 | -.317 | .256 | -.032 | .088 | -.093 | .133 | -.005 | **.689** |

Table 3: Principal Components Analysis

| Distressed Communities Factor | |
|---|---|
| Rank | Communities |
| 1 | Kashmere Gardens |
| 2 | Greater Fifth Ward |
| 3 | Westwood |
| 4 | Greater Third Ward |
| 5 | OST/ South Union |
| 6 | Independence Heights |
| 7 | Settegast |
| 8 | Gulfton |
| 9 | Greater Greenspoint |
| 10 | SunnySide |
| | |
| 79 | Washington Avenue/ Memorial Park |
| 80 | Clear Lake |
| 81 | Braeswood Place |
| 82 | Greater Uptown |
| 83 | Greenway/ Upper Kirby Area |
| 84 | Kingwood Area |
| 85 | Memorial |
| 86 | Lake Houston |
| 87 | University Place |
| 88 | Afton Oaks/ River Oaks Area |

Table 4: Community Ranks on Distressed Communities Factor

Communities Ranked 1 – 10 will be considered Group 1, which are the top ten most Distressed communities in Houston. Communities 79 – 88 are Group 2, which is the top ten least distressed (Wealthy) communities in Houston.

*Two-Independent Samples-T-Test*
The next step is to run a Two-Independent Sample T-Test to evaluate difference between the two groups identified above based on the measures that relate to public investment. These are measures, which are directly affected by the level of local government investment and development. Those measures are as follows.

| Public Investment Indicators | |
|---|---|
| VMT | Percent Housing Units close to Business Centers |
| Population ¼ mile to transit | Parks Access |
| Percent of Open Space | Population Close to Waste Sites |
| Intersection Density | Percent Low-Mid Developed Land Use |
| Food Deserts | Street Condition |
| Percent High Developed Land Use | Population in flood zone |
| Distance to CBD | Storm Sewer condition |

The following table shows group statistics for our two groups. We can see clear differences between the means for some of the indicators such as: Population ¼ mile to transit, Intersection Density, Percent High Developed Land Use, Distance to CBD, Percent Housing Units close to Business Centers, Population Close to Waste Sites, Storm Sewer condition.

Group Statistics: Group 1 – Distressed Communities; Group 2 – Wealthy Communities

| | Distressed | N | Mean | Std. Deviation | Std. Error Mean |
|---|---|---|---|---|---|
| VMT | 1.00 | 10 | 17,294.844 | 2,100.947 | 664.378 |
| | 2.00 | 10 | 17,472.863 | 5,199.056 | 1,644.086 |
| Population ¼ mile to transit | 1.00 | 10 | 83.282 | 14.470 | 4.576 |
| | 2.00 | 10 | 57.496 | 39.409 | 12.462 |
| Percent of Open Space | 1.00 | 10 | 10.402 | 8.464 | 2.677 |
| | 2.00 | 10 | 10.703 | 5.405 | 1.709 |
| Intersection Density | 1.00 | 10 | 103.075 | 61.878 | 19.568 |
| | 2.00 | 10 | 75.66 | 29.667 | 9.381 |
| Food Deserts Population | 1.00 | 10 | 41.634 | 29.837 | 9.435 |
| | 2.00 | 10 | 41.999 | 37.022 | 11.7076 |
| Percent High Developed Land Use | 1.00 | 10 | 27.699 | 18.802 | 5.946 |
| | 2.00 | 10 | 18.600 | 14.987 | 4.739 |
| Distance to CBD | 1.00 | 10 | 6.478 | 3.87 | 1.224 |
| | 2.00 | 10 | 10.462 | 7.382 | 2.334 |
| Percent Housing Units close to Business Centers | 1.00 | 10 | 20.88 | 27.414 | 8.669 |
| | 2.00 | 10 | 49.56 | 41.770 | 13.209 |
| Parks Access | 1.00 | 10 | 40.27 | 15.220 | 4.813 |
| | 2.00 | 10 | 46.46 | 22.100 | 6.989 |
| Population Close to Waste Sites | 1.00 | 10 | 1,177.72 | 1,888.960 | 597.342 |
| | 2.00 | 10 | 521.33 | 559.901 | 177.056 |
| Percent Low-Mid Developed Land Use | 1.00 | 10 | 58.463 | 16.975 | 5.368 |
| | 2.00 | 10 | 51.829 | 20.931 | 6.619 |
| Street Condition | 1.00 | 10 | 17.91 | 8.066 | 2.551 |
| | 2.00 | 10 | 22.28 | 13.313 | 4.210 |
| Population in flood zone | 1.00 | 10 | 22.716 | 23.592 | 7.460 |
| | 2.00 | 10 | 23.523 | 25.377 | 8.025 |
| Storm Sewer condition | 1.00 | 10 | 13.918 | 10.876 | 3.439 |
| | 2.00 | 10 | 29.157 | 24.144 | 7.635 |

Our next step is to decide if the mean differences (Table x) are large enough to be considered significant differences between the Distressed communities and the Wealthy communities. Our values were standardized since the sample sizes were small. Using the Levene's Test for equality of variances, we find that five of the 14 measures do not have equal variances. Of these five, only the *'Percent of the population within ¼ mile of transit'* variable, shows a significant difference between the Distressed communities and Wealthy communities. The table below shows that of the fourteen (14) public investment variables, only three (3) show significant differences between the Distressed communities and the Wealthy communities. Those variables are: Population ¼ mile to transit, Percent Housing Units close to Business Centers, and Storm Sewer condition.

| Equality of Variances | t | df | Sig. (2-tailed) | Mean Difference | Std. Error Difference | 95% Confidence Interval of the Difference |
|---|---|---|---|---|---|---|

|  | F | Sig. |  |  |  |  |  | Lower | Upper |
|---|---|---|---|---|---|---|---|---|---|
| VMT | 9.668 | 0.006 | -0.1 | 18 | 0.921 | -0.057 | 0.567 | -1.248 | 1.134 |
| **Population ¼ mile to transit** | **8.58** | **0.009** | **8.169** | **18** | **0** | **2.37** | **0.29** | **1.76** | **2.978** |
| Percent of Open Space | 2.026 | 0.172 | -0.095 | 18 | 0.926 | -0.038 | 0.402 | -0.883 | 0.807 |
| Intersection Density | 3.872 | 0.065 | 1.263 | 18 | 0.223 | 0.464 | 0.367 | -0.307 | 1.234 |
| Food Deserts Population | 1.163 | 0.295 | -0.024 | 18 | 0.981 | -0.011 | 0.464 | -0.987 | 0.964 |
| Percent High Developed Land Use | 0.276 | 0.606 | 1.197 | 18 | 0.247 | 0.601 | 0.502 | -0.454 | 1.656 |
| Distance to CBD | 9.77 | 0.006 | -1.511 | 18 | 0.148 | -0.858 | 0.568 | -2.052 | 0.335 |
| **Percent Housing Units close to Business Centers** | **3.105** | **0.095** | **-1.816** | **18** | **0.086** | **-0.834** | **0.459** | **-1.8** | **0.131** |
| Parks Access | 1.865 | 0.189 | -0.729 | 18 | 0.475 | -0.282 | 0.387 | -1.095 | 0.531 |
| Population Close to Waste Sites | 6.128 | 0.023 | 1.054 | 18 | 0.306 | 0.6 | 0.57 | -0.597 | 1.797 |
| Percent Low-Mid Developed Land Use | 0.531 | 0.476 | 0.779 | 18 | 0.446 | 0.411 | 0.528 | -0.698 | 1.52 |
| Street Condition | 1.863 | 0.189 | -0.888 | 18 | 0.386 | -0.324 | 0.364 | -1.09 | 0.442 |
| Population in flood zone | 0.228 | 0.639 | -0.074 | 18 | 0.942 | -0.04 | 0.547 | -1.19 | 1.109 |
| **Storm Sewer condition** | **6.737** | **0.018** | **-1.82** | **18** | **0.085** | **-0.75** | **0.412** | **-1.616** | **0.116** |

## DISCUSSION

*Population living ¼ mile to transit*

**Group Statistics: Group 1 – Distressed Communities; Group 2 – Wealthy Communities**

|  | Distressed | N | Mean | Std. Deviation | Std. Error Mean |
|---|---|---|---|---|---|
| Population ¼ mile to transit | 1.00 | 10 | 83.282 | 14.470 | 4.576 |
|  | 2.00 | 10 | 57.496 | 39.409 | 12.462 |

The Metropolitan Transit Agency in Houston (METRO) provides a good service of public transportation to the Distressed communities. On average 83% of the people in Distressed communities live within ¼ mile of a transit stop (walking distance). It is not clear that opportunities exist to increase the proximity of persons in the Wealthy communities to transit stops. Or for that matter, if this increased accessibility would improve the transit ridership among that community.

*Percent Housing Units close to Business Centers*

**Group Statistics: Group 1 – Distressed Communities; Group 2 – Wealthy Communities**

|  | Distressed | N | Mean | Std. Deviation | Std. Error Mean |
|---|---|---|---|---|---|
| Percent Housing Units close to Business Centers | 1.00 | 10 | 20.88 | 27.414 | 8.669 |
|  | 2.00 | 10 | 49.56 | 41.770 | 13.209 |

This indicator demonstrates the opportunity and need to increase affordable housing units close to business centers. It may also signal the need to increase economic development opportunities close to Distressed communities. Houston has approximately 14 business centers located at various diverse locations across the 660sqml city. The City has historically adopted a laissez faire approach to economic development and as a result the business community has led the determination of ideal location for businesses. Cities like Houston, can do more to encourage the preferred location of businesses by adding incentives to businesses and by investing in infrastructure necessary to improve business efficiency in locations of choice. Infrastructure such as high-speed communication networks; special district designation for improvement management efficiency go a long way in incentivizing business location decisions.

*Storm Sewer condition*

**Group Statistics: Group 1 – Distressed Communities; Group 2 – Wealthy Communities**

|  | Distressed | N | Mean | Std. Deviation | Std. Error Mean |
|---|---|---|---|---|---|
| Storm Sewer condition | 1.00 | 10 | 13.918 | 10.876 | 3.439 |
|  | 2.00 | 10 | 29.157 | 24.144 | 7.635 |

The storm sewer condition between the Distressed communities and the Wealthy communities in Houston are significantly different. On average 14% of the storm sewers in the Distressed communities are adequate. While on average 29% of the storm sewers in Wealthy communities are adequate. Both groups have relatively low numbers, however these results show that local government should prioritize storm sewer improvements in the Distressed communities over those in the Wealthy communities to reach more of a state of equity in the provision of this public service.

**CONCLUSION**

Public Choice theory describes trends in Houston, it suggests that all things should be subject to market forces and should not be distorted by government regulation. The problem is that people are not rational utility maximizers and hence there is an imbalance between big business representation and citizen interests (Keating 1995). Zoning is seen as opposing to public choice theory and some planners also prefer not to have zoning since it has become too restrictive in places hoping to implement mixed use development. Robert Litke, former planning director of Houston, states that Houston has developed much like other Metro places therefore land use regulations may not be necessary (Neuman 2003). Proponents of zoning of course highlight many valid positions including anticipating development decisions; protecting fragile land and providing social balance in terms of protecting public property from market prospectus (Keating 1995). Still another opinion is that zoning may be an outdated tool since it is restrictive but we need something to protect fragile landscapes and the public from the excesses of the market. This last opinion is more in favor of comprehensive planning to guide decision making (Berger 2004).

Urban planning has matured into a very complex system of management of social, physical, institutional and policy environments. The diverse nature of these environments must be reduced or integrated in a way in which to facilitate change. In our society the three competing interests of social, market and environmental perspectives, each have a stake in these changing diverse environments and as such should be the gauge in which to balance the environments. Using social, market and environmental perspectives as the gauge for balance we can apply community discourse to achieve that goal. However we should keep in mind that the process itself would be dependent on the distribution of power and organized interests (Kaiser, Godschalk, & Chapin, 1995).

Indicators are used to serve political and operational functions. They support problem recognition and awareness; communication; opinion forming and strategic design for problem solving. On the operational gamut they are useful for goal oriented objectives, as well as monitoring, control and reporting (Backhaus, Bock, & Weiers, 2002). Indicators can be characterized as having intrinsic and extrinsic value. The intrinsic value can be defined as characterization of the element being measured. The extrinsic value appears where indicators may also report characteristics about other systems or other elements in an environment (Bossel, 1999). Indicators can serve to explain, predict, justify and offer normative guidance on the achievement of goals for the development of human settlements (Neuman, 2005). This speaks to their exceptional importance in supporting the adoption of successful smart growth programs.

References

U.S. Census Bureau. (2011, December). American FactFinder. Washington, DC. Retrieved from http://www.census.gov

Altshuler, A. (1965). The goals of comprehensive planning. *Journal of the American Institute of Planners, 31*(3), 186-195.

APA. (2010). *APA's Sustainaing Places Initative*. Retrieved October 2013, from American Planning Association: https://www.planning.org/sustainingplaces/

APA. (2012). *Comprehensive plan standards for sustaining places*. Retrieved October 2013, from http://www.planning.org/sustainingplaces/compplanstandards/

ASCE. (2012). *2012 Report Card for Houston Area Infrastructure.* Houston, TX: American Society of Civil Engineering - Texas Section Houston Branch.

Berke, P., & Godschalk, D. (2009). Searching for the good plan: A meta-analysis of plan quality studies. *Journal of Planning Literature, 23*(3), 227-240.

Berke, P., & Manta-Conroy, M. (2000). Are we Planning for Sustainable Development? An Evaluation of 30 Comprehensive Plans. *Journal of the American Planning Association*, 21-33.

BISNOW. (2013, November). *Grocery Wars*. Retrieved November 2013, from Real Estate Bisnow: http://www.bisnow.com/commercial-real-estate/houston/grocery-wars/

Brennan, M. (2012, July 26). Houston tops our list of america's coolest cities. *Forbes*, p. 2.

Burchell, R., Downs, A., McCann, B., & Mukherji, S. (2005). *Sprawl Costs: Economic Impacts of Unchecked Development.* Washington: Island Press.

Campbell, S. (1996). Green cities, growing cities, just cities? Urban planning and the contradictions of sustainable development. *Journal of the American Planning Association, 62*(3), 296-312.

Centers for Disease Control and Prevention. (2012). *A Look Inside Food Deserts*. Retrieved November 2012, from www.cdc.gov/features/fooddeserts

Conroy, M. M., & Beatley, T. (2007). Getting it done: An exploration of sustainability efforts in practice. *Planning, Practice & Research, 22*(1), 25-40.

Economist. (2009, July). The new face of America: Texas is the bellwether for demographic change across the country. *The Economist*.

Elgert, L., & Krueger, R. (2012). Modernising sustainable development? Standardisation, evidence and experts in local indicators. *Local Environment: The International Journal of Justice and Sustainability, 17*(5), 561-571.

Emmett, E. (2013, April). Harris County Judge Ed Emett delivers a talk at the Kinder Institute Annual Luncheon. Houston, TX: Kinder Institute.

Fainstein, S. (2000). New Directions in Planning Theory. *Urban Affairs Review, 35*(4), 451-478.

Gafrick, M. (2013). Role of special districts in successfully implementing community plans. *TxAPA Annual Conference.* Galveston, TX: American Planning Association Texas Chapter.

Godschalk, D., & Anderson, W. (2012). *Sustaining places: The role of the comprehensive plan.* APA PAS Report.

Hatry, H. (2006). *Performance measurment: getting results* (2 ed.). Washington, DC: Urban Institute Press.

HNTB, & Spillette, S. (2003). *Compendium of plans*. Houston, TX: BlueprintHouston.

Innes, J. (1988). The power of data requirements. *Journal of the American Planning Association, 54*(3), 275-278.

Jepson, E., & Haines, A. (2014). Zoning for sustainability: A review and analysis of the zoning ordinances of 32 cities in the United States. *Journal of the American Planning Association, 80*(3), 239-252.

Kaiser, E. J., & Godschalk, D. (1995). Twentieth century land use planning: A stalwart family tree. *Journal of the American Planning Association, 61*(3), 365-385.

Kaiser, E. J., Godschalk, D. R., & Chapin, F. S. (1995). *Urban Land Use Planning* (Vol. 4). Chicago, IL: University of Illinois Press.

Kaza, N., & Hopkins, L. (2007). Ontology for land development decisions and plans. Studies in computational intelligence. In J. Teller, J. Lee, & C. Roussey (Eds.), *Ontologies for urban development* (pp. 47-59). Berlin: Springer-Verlag.

Kaza, N., & Hopkins, L. (2012). Intentional actions, plans, and information systems. *International Journal of Geographical Information Science, 26*(3), 557-576.

King, L. (2012). *Houston Sustainable Development Indicators: A Comprehensive Development Review for Citizens, Analysts and Decision Makers*. Houston: Shell Center for Sustainability, Rice University.

King, L. (2013). *Sustainable development of Houston districts: The health of the City*. Houston, TX: Shell Center for Sustainability.

Levy, J. (1992). What has happened to planning? *Journal of the American Planning Association, 58*(1), 81-84.

Mori, K., & Christodoulou, A. (2012). Review of sustianability indices and indicators: Toward a new city sustainability index (CSI). *Environmental Impact Assessment Review, 32*(1), 94-106.

Myers, D. (1997). Anchor Points for Planning's Identification. *Journal of Planning Education and Research*, 223-224.

Neuman, M. (1998). Does Planning Need the Plan? *Journal of the American Planning Association, 64*(2), 208-220.

Neuman, M. (2005). Notes on the Uses and Scope of City Planning Theory. *Planning Theory, 4*(2), 123-145.

Neuman, M., & Churchill, S. W. (2011). A General Process Model of Sustainability. *Industrial and Engineering Chemistry Research*, 8901-8904.

Nijkamp, P., & Vreeker, R. (2000). Sustainability assessment of development scenarios: Methodology and application to Thailand. *Ecological Economics, 33*, 7-27.

Quay, R. (2010). Anticipatory Governance. *Journal of the American Planning Association, 76*(4), 496-511.

Rudick, T. (2013, September 25). Houston's not so affordable anymore: New study ranks it on 26th among major U.S. cities. *CultureMap*, p. 1.

SANDAG. (2006). *Regional comprehensive plan performance monitoring*. Retrieved October 2013, from Comprehensive Land Use and Regional Growth Projects: http://www.sandag.org/index.asp?projectid=309&fuseaction=projects.detail

Sarnoff, N. (2013, September 24). Housing costs put hurt on incomes. *Houston Chronicle*, p. 1.

SCS. (2013). *Houston Sustainability Indicators*. Retrieved November 2013, from

https://shellcenter.rice.edu/Content.aspx?id=2147483977

Talen, E. (1996). Do plans get implemented? A review of evaluation in planning. *Journal of Planning Literature, 10*(3), 248-259.

UN. (1993). *Report of the united nations conference on environment and development.* Washington, DC: United Nationa.

UN. (2007). *Indicators of Sustainable Development: Guidelines and Methodologies, 3rd Ed.* New York, NY: United Nations.

UNCED. (1992). *Agenda 21: Programme of action for sustainable development.* Washington, DC: United Nations Department of Public Information.

United Nations. (1996). *Indicators of Sustainable Development Framework and Methodologies.* New York, NY: United Nations.

United States Congress. (1993, August 3). Government performance and results act of 1993. *107 STAT, 285*, Public Law 103-62. Washington, DC.

US Census. (2007). *County and City Data Book 2007.* Washington, DC: US Census Bureau.

US Green Building Council. (2013). *LEED in motion: People and progress.* Washington, DC: US Green Building Council.

USGBC. (2009). *LEED for Neighborhood Development*. Retrieved December 2014, from http://www.usgbc.org/resources/leed-neighborhood-development-v2009-current-version

Warner, M. (2011). Club Goods and Local Government. *Journal of the American Planning Association, 77*(2), 155-166.

WCED. (1987). *Our Common Future.* Oxford, UK: Oxford University Press.

WHO. (2013). *Urban population growth*. Retrieved December 2013, from http://www.who.int/gho/urban_health/situation_trends/urban_population_growth_text/en/

Wiek, A., & Binder, C. (2005). Solution spaces for decision-making: a sustainability assessment tool for city-regions. *Environmental Impact Assessment Review, 25*, 589-608.